\documentclass[8.5pt,twoside,twocolumn]{article}
\usepackage[super,sort&compress,comma]{natbib}
\usepackage{mhchem}
\usepackage{times,mathptmx}
\usepackage{sectsty}
\usepackage{balance}

\usepackage{graphicx} %eps figures can be used instead
\usepackage{lastpage}
\usepackage[format=plain,justification=raggedright,singlelinecheck=false,font=small,labelfont=bf,labelsep=space]{caption}
\usepackage{fancyhdr}
\usepackage{color}

\begin{document}

\thispagestyle{plain}
\fancypagestyle{plain}{
% \fancyhead[L]{\includegraphics[height=8pt]{headers/LH}}
% \fancyhead[C]{\hspace{-1cm}\includegraphics[height=20pt]{headers/CH}}
% \fancyhead[R]{\includegraphics[height=10pt]{headers/RH}\vspace{-0.2cm}}
\renewcommand{\headrulewidth}{1pt}}
\renewcommand{\thefootnote}{\fnsymbol{footnote}}
\renewcommand\footnoterule{\vspace*{1pt}%
\hrule width 3.4in height 0.4pt \vspace*{5pt}}
\setcounter{secnumdepth}{5}

\makeatletter
\def\subsubsection{\@startsection{subsubsection}{3}{10pt}{-1.25ex plus -1ex minus -.1ex}{0ex plus 0ex}{\normalsize\bf}}
\def\paragraph{\@startsection{paragraph}{4}{10pt}{-1.25ex plus -1ex minus -.1ex}{0ex plus 0ex}{\normalsize\textit}}
\renewcommand\@biblabel[1]{#1}
\renewcommand\@makefntext[1]%
{\noindent\makebox[0pt][r]{\@thefnmark\,}#1}
\makeatother
\renewcommand{\figurename}{\small{Fig.}~}
\sectionfont{\large}
\subsectionfont{\normalsize}

\fancyfoot{}
% \fancyfoot[LO,RE]{\vspace{-7pt}\includegraphics[height=9pt]{headers/LF}}
% \fancyfoot[CO]{\vspace{-7.2pt}\hspace{12.2cm}\includegraphics{headers/RF}}
% \fancyfoot[CE]{\vspace{-7.5pt}\hspace{-13.5cm}\includegraphics{headers/RF}}
\fancyfoot[RO]{\footnotesize{\sffamily{1--\pageref{LastPage} ~\textbar  \hspace{2pt}\thepage}}}
\fancyfoot[LE]{\footnotesize{\sffamily{\thepage~\textbar\hspace{3.45cm} 1--\pageref{LastPage}}}}
\fancyhead{}
\renewcommand{\headrulewidth}{1pt}
\renewcommand{\footrulewidth}{1pt}
\setlength{\arrayrulewidth}{1pt}
\setlength{\columnsep}{6.5mm}
\setlength\bibsep{1pt}

\newcommand{\alt}{\raisebox{-0.3ex}{$\stackrel{<}{\sim}$}}
\newcommand{\agt}{\raisebox{-0.3ex}{$\stackrel{>}{\sim}$}}

\twocolumn[
  \begin{@twocolumnfalse}
\noindent\LARGE{\textbf{A quantum chemical study from a molecular perspective: ionization and electron attachment energies for species often used to fabricate single-molecule junctions
}}
\vspace{0.6cm}

\noindent\large{\textbf{Ioan B\^aldea $^{\ast}$
\textit{$^{a\ddag}$}
}}\vspace{0.5cm}

% \noindent\textit{\small{\textbf{Received Xth XXXXXXXXXX 20XX, Accepted Xth XXXXXXXXX 20XX\newline
% First published on the web Xth XXXXXXXXXX 200X}}}

\noindent \textbf{\small{Published: Faraday Discussions 2014, {\bf 174}, 37-56; DOI: 10.1039/C4FD00101J}}
\vspace{0.6cm}

\noindent 
\normalsize{Abstract:\\
The accurate determination of the lowest electron attachment ($EA$) and ionization ($IP$)
energies for molecules embedded in molecular junctions is important
for correctly estimating, \emph{e.g.}, the magnitude of the currents ($I$) or 
the biases ($V$) where an $I-V$-curve exhibits a significant non-Ohmic behavior.  
Benchmark calculations for the lowest electron attachment and ionization energies of 
several typical molecules utilized to fabricate single-molecule junctions characterized by n-type
conduction (4,4'-bipyridine, 1,4-dicyanobenzene, and 4,4'-dicyano-1,1'-biphenyl) and p-type conduction
(benzenedithiol, biphenyldithiol, hexanemonothiol, and hexanedithiol] 
based on the EOM-CCSD (equation-of-motion coupled-cluster singles and doubles) state-of-the-art 
method of quantum chemistry are presented.
They indicate significant 
differences from the results obtained within current approaches to molecular  
transport. 
The present study emphasizes that, in addition to a reliable quantum chemical method,
basis sets much better than the ubiquitous double-zeta set employed 
for transport calculations are needed. The latter is a particularly critical issue
for correctly determining $EA$'s, which is impossible without including sufficient 
diffuse basis functions. The spatial distribution of the dominant molecular
orbitals (MO's) is another important issue, on which the present study draws attention,
because it sensitively affects the MO-energy shifts $\Phi$ due to image charges 
formed in electrodes. The present results cannot substantiate 
the common assumption of a point-like MO midway between electrodes,
which substantially affects the actual $\Phi$-values.

$ $ \\  

{{\bf Keywords}: molecular electronics, 
% charge transport, 
single-molecule junctions, quantum chemical calculations, electron attachment, ionization}
}
\vspace{0.5cm}
 \end{@twocolumnfalse}
  ]

%Footnotes
%Please use \dag to cite the ESI in the main text of the article.
%If you article does not have ESI please remove the the \dag symbol from the title and the above footnotetext.

\footnotetext{\textit{$^{a}$~Theoretische Chemie, Universit\"at Heidelberg, Im Neuenheimer Feld 229, D-69120 Heidelberg, Germany.}}
\footnotetext{\ddag~E-mail: ioan.baldea@pci.uni-heidelberg.de.
Also at National Institute for Lasers, Plasmas, and Radiation Physics, Institute of Space Sciences,
Bucharest, Romania}
\section{Introduction}
\label{sec:introd}
Electron or hole injection into molecules embedded between two electrodes 
represents an important issue in the fabrication of molecular devices. 
The efficiency of the charge injection and transport in molecular junctions is 
controlled by the highest occupied or lowest unoccupied molecular orbital (HOMO or LUMO, respectively),
whichever is closest to the electrodes' Fermi energy $E_F$, and the key quantity is 
the energy offset $\varepsilon_0 = \min(E_F - E_{HOMO}, E_{LUMO} - E_F)$.
It can be compared to a tunneling barrier, which charge carriers have to overcome to generate a current. 
Ultraviolet photoelectron spectroscopy (UPS) \cite{Frisbie:11}, thermopower \cite{Reddy:08,Venkataraman:12a,Tao:13},
and transition voltage spectroscopy (TVS) \cite{Beebe:06,Araidai:10,Baldea:2010h,Guo:11,Baldea:2012g,Vuillaume:12c}
represent current methods to estimate the relative alignment of the dominant molecular orbital from experimental
data. 

Recent analysis of a variety of transport data 
demonstrated that full current-voltage $I-V$ curves beyond the ohmic regime can 
be quantitatively reproduced by assuming that molecular transport is dominated by a single 
level (to be identified with HOMO or LUMO) 
% \cite{Baldea:2010h,Baldea:2012a,Baldea:2012b,Baldea:2012g,Baldea:2013b,Baldea:2013d,Baldea:2014a}.
\cite{Baldea:2010h,Baldea:2012a,Baldea:2012b,Baldea:2012g,Baldea:2013b,Baldea:2013d}.
This is an enormous simplification. Still, the correct description of the relative alignment 
$\varepsilon_0$ ($=E_{LUMO} - E_F$ or $E_F - E_{HOMO}$) remains an important challenge for \emph{ab initio} 
approaches to the charge transport through single-molecule junctions. 
It is a challenge particularly because of the high accuracy needed. 
Values directly determined in ultraviolet photoelectron spectroscopy (UPS) experiments
amount to $\varepsilon_0 \sim 1$\,eV \cite{Frisbie:11}. Results based on TVS by using a model 
able to excellently reproduce current-voltage ($I-V$) curves measured in single-molecule junctions
demonstrate that $\varepsilon_0$ can be even smaller ($\varepsilon_0 \simeq 0.6$\,eV \cite{Baldea:2013b}).
It should be clear that, in view of such low-$\varepsilon_0$ values, estimates for $IP$'s and $EA$'s 
with errors $\simeq $0.5\,eV typical for quantum chemical methods of moderate accuracy are unacceptable.
Noteworthy, the quantity $\varepsilon_0$ is important not only because it determines the magnitude of the currents,
but also because it indicates the biases beyond which an $I-V$ curve 
becomes significantly nonlinear \cite{Baldea:2012b}.

An important message, which the present study aims to convey, is that 
the accurate determination of the 
HOMO and LUMO energies (or, more precisely, the lowest ionization $IP$ and electroaffinity $EA$
with reversed sign) represent a nontrivial issue even for isolated species 
of interest for molecular transport. Obviously, this is a minimal requirement for any 
molecular transport approach.

In the present paper we report results of ongoing work 
focusing on several prototypical molecular species, which are mostly utilized in the 
fabrication of single-molecule junctions, and examine the reliability of the results for 
ionization energies $IP$($\approx -E_{HOMO}$) and 
electroaffinities $EA$($\approx -E_{LUMO}$) as obtained within current methods
employed in molecular transport with accurate estimates 
obtained within well-established quantum chemical methods.
To avoid misunderstandings, let us explicitly mention that 
we will restrict ourselves throughout to the lowest $IP$ and $EA$; in fact,
if at all, for the molecules analyzed below 
it is only the lowest $EA$ that corresponds to a stable anion ($EA > 0$).

The molecular species to be considered 
will include, besides 4,4'-bipyridine (44BPY) 
--- a molecule recently considered from a TVS perspective 
% \cite{Baldea:2012i,Baldea:2013a,Baldea:2013b,Baldea:2013c,Baldea:2014a} ---, 
\cite{Baldea:2013b,Baldea:2013c} ---, 
molecules often used to fabricate two classes of molecular junctions.
One class is represented by molecules characterized by n-type (LUMO-mediated) conduction,
the other class 
comprises molecules characterized by p-type (HOMO-mediated) conduction.
As specific examples belonging to the first class, we will consider 
44BPY \cite{Venkataraman:12a}, 
BDCN (1,4-dicyanobenzene) \cite{Reed:09}, and 2BDCN (4,4'-dicyano-1,1'-biphenyl).
From the second class, we will examine oligophenylene dithiols \cite{Guo:11,Tao:13}
and alkanemono- and dithiols \cite{Reed:11}.

Because of well-documented shortcomings of ubiquitous methods based on 
density functional theory (DFT), emphasis will be on post-DFT methods. 
Besides methods already utilized in approaches to molecular transport 
(e.g., $\Delta$-SCF \cite{Gunnarson:89}, $GW$ \cite{Gunnarson:98} and MP2 \cite{Shimazaki:06b}), 
we will consider truly \emph{ab initio} methods 
used in quantum chemistry: outer valence Green's functions (OVGF) \cite{Cederbaum:75,Schirmer:84}, 
second-order algrabraic-diagrammatic constructions [ADC(2)] \cite{Schirmer:82,Schirmer:91}, and
equation-of-motion (EOM) coupled-cluster (CC) \cite{Stanton:93} approaches. 
The importance of using appropriate quantum chemical methods 
will be emphasized. As is well known, calculating $EA$'s 
is a very delicate problem in quantum chemistry altogether \cite{Klaiman:13}.
%%%%%%%%%%%%%%%%%%%%%%%%%%%%%%%%%%%%%%%%%%%%%%%%%%%%%%%%%%%%%%%%%%%%%%%%%%%%%%%% 
\section{Methods}
\label{sec:methods}
%%%%%%%%%%%%%%%%%%%%%%%%%%%%%%%%%%%%%%%%%%%%%%%%%%%%%%%%%%%%%%%%%%%%%%%%%%%%%%%%
Active molecules embedded in molecular junctions can be treated at various levels 
of theory ranging from tight-binding (extended H\"uckel) to and post-DFT.
In order to facilitate understanding the message, which the results reported below
aim to convey, we will briefly present the methods utilized in this study.

(i) If the picture based on the \emph{self-consistent field} (SCF) were valid 
(or, equivalently, electron correlations were absent), 
the energy of highest occupied Hartree-Fock (HF) orbital (i.e., HOMO) with reversed sign
would represent the lowest ionization energy $IP = - E_{SCF, HOMO}$ (Koopmans theorem). 
The physical meaning of the virtual (unoccupied) is controversial \cite{Schulman:67,Hunt:69}. 
Virtual HF orbitals might have physical meaning if descriptions based on small basis sets
succeeded (at least semi-)quantitatively,
but this is often not the case. Table \ref{table-bdt-mp2} 
illustrates this failure, where results relevant in connection with existing
MP2-based transport approaches and minimal (STO-6G) basis sets \cite{Shimazaki:09} are presented. 
At the other extreme, it is also well known that the HF LUMO energy
goes to zero in the complete basis set limit \cite{Garza:00,Musgrave:06}.
This is a reason why attempts to ``improve'' the quality of a transport approach
by using larger and larger basis sets end up with unphysical results \cite{Ratner:10}. 
For large atomic orbital (AO) basis sets, the virtual HF orbitals
have mathematical rather than physical meaning, namely, in providing
an expansion manifold for the physical states of interest.
In some cases, individual virtual or unoccupied HF orbitals 
(in particular, the LUMO) can reasonably describe, e.g.,
anionic bound or resonance states semi-quantitatively, provided
that the size of the AO basis used is not too large \cite{Baldea:2013b}. 

(ii) Within \emph{DFT-approaches}, the single particle solutions 
of the Kohn-Sham (KS) equations are handled as if the corresponding 
eigenvalues/eigenfunctions were real orbital energies/wave functions.
The corresponding implementation in a Landauer-NEGF formalism is straightforward
because the DFT description is mathematically a single-particle description. 
This is why the DFT approaches to molecular transport 
are by far the most popular to date.
Drawbacks of such DFT-approaches are well documented. The drastic underestimation 
of the HOMO-LUMO gap and the related lineup problem 
(HOMO/LUMO energies too close to electrodes' Fermi energy) are issues most frequently 
mentioned and not at all surprising:
as is well known \cite{Gunnarson:89}, KS ``orbitals'' are mathematical objects rather
than physical orbitals.\\

The quantum chemical methods used in the present paper to compute the lowest electroaffinity ($EA$) 
and ionization $IP$ energies are:\\

(iii) The \emph{outer valence Green's function (OVGF) method}
\cite{Cederbaum:77,Schirmer:84} represents the most elaborate quantum chemical approach 
based on a single-particle picture. To the best of our knowledge, the
OVFG approach has not yet been
utilized in molecular transport studies. Therefore, let us mention that
the OVGF method is a way to approximately include the contribution
of the electron-electron interaction beyond HF. 
Details can be found in ref.~\cite{Cederbaum:77,Schirmer:84}.
As explained below, this method is superior to the MP2-like approximation
used recently \cite{Shimazaki:06b} (see sec.~\ref{sec:basis}).
The {OVGF method \cite{Cederbaum:77,Schirmer:84} exactly treats the full 
second- and third-order terms
in the self-energy entering the Dyson equation for the one-electron Green function,
and is augmented by a geometrical approximation to
also include further higher-order corrections \cite{dyson}.
Results for the electron affinities and ionization energies
obtained by considering the second- and third-order terms
are shown in the tables presented below 
(they are labeled as ``2nd-order pole'' and ``3rd-order pole'', respectively) 
along with those of the full OVGF.
The corresponding pole strengths \cite{Schirmer:84}
are also indicated (percents in parentheses).

(iv) \emph{The algebraic-diagrammatic construction (ADC)} is based on a 
diagrammatic perturbation expansion. ADC(n) defines an approximate scheme 
of infinite partial summations exact up to the
$n$-th order of perturbation theory \cite{Schirmer:82,Schirmer:91}.
Its second-order version ADC(2) is superior to the so-called $GW$-approximation
\cite{Gunnarson:98}, since only bubble contributions are included within the $GW$
and not all second-order terms:
expressing the self-energy $\boldsymbol{\Sigma}$ by the product of the single-electron 
Green's function $\boldsymbol{G}$ and the effective interaction $\boldsymbol{W}$ 
($ \boldsymbol{\Sigma} \sim \boldsymbol{G W}$, thence the name $GW$), 
vertex corrections \cite{mahan} are neglected within $GW$.  

(iv) \emph{Equation-of-motion (EOM) coupled-cluster (CC) approaches} 
at singles and doubles (CCSD) \cite{Nooijen:95a}, hybrid (CC2) \cite{Christiansen:95}, and perturbative [CCSD(2)]
\cite{Stanton:95,Nooijen:95b} levels will be extensively applied in this paper. 
Corrections due to triples [EOM-CCSD(T)] will be also considered; 
the fact that these corrections are altogether negligible
is an indication on the accuracy of the state-of-the-art EOM-CCSD method.

(vi) \emph{Energy difference ($\Delta$-)} methods will be utilized for all the aforementioned 
cases. Within these methods, 
the lowest ionization energy $IP$ and electron affinity $EA$  
are estimated as differences between the ground  state energies $\mathcal{E}$
of the corresponding molecular charge species ($M$=SCF, DFT, MP2, CCSD, CC2)
at the equilibrium geometry of the neutral molecule
%%%%%%%%%%%%%%%%%%%%%%%%%%%%%%%%%%%%%%%%%%%%%%%%%%%%%%%%%%%%%%%%%%%%%%%%%%%%%%%%
\begin{eqnarray}
\label{eq-delta-i}
IP \to \Delta_{M}^{i} & = & \mathcal{E}_{M, cation} - \mathcal{E}_{M, neutral} , \\
EA \to \Delta_{M}^{a} & = & \mathcal{E}_{M, neutral} - \mathcal{E}_{M, anion} .
\label{eq-delta-a}
\end{eqnarray}
%%%%%%%%%%%%%%%%%%%%%%%%%%%%%%%%%%%%%%%%%%%%%%%%%%%%%%%%%%%%%%%%%%%%%%%%%%%%%%%%
%%%%%%%%%%%%%%%%%%%%%%%%%%%%%%%%%%%%%%%%%%%%%%%%%%%%%%%%%%%%%%%%%%%%%%%%%%%%%%%% 
\section{Computational details}
\label{sec:details}
%%%%%%%%%%%%%%%%%%%%%%%%%%%%%%%%%%%%%%%%%%%%%%%%%%%%%%%%%%%%%%%%%%%%%%%%%%%%%%%%
The results of the SCF, DFT/B3LYP, MP2, and OVGF calculations reported below were done with  
GAUSSIAN 09 \cite{g09}. Coupled-cluster calculations of the $IP$, $EA$, 
total energies of the various charge species and excitation energies 
were performed with CFOUR \cite{cfour}.
Calculations within the so-called regular (strict) ADC(2) reported here
have been done with the fully parallelized PRICD-$\Sigma$(2) code
\cite{Vysotskiy:10a}, which is interfaced to MOLCAS \cite{molcas}. As amply documented
by extensive work of the Heidelberg theoretical chemistry group, the results 
based on the strict ADC(2) are comparable to the
second-order approximate coupled cluster singles and doubles model (CC2) \cite{Christiansen:95}.
Augmented with extra terms
% to employ the first-order expansion of the matrix elements of the highest
% block (for EA calculations, this is the $3h-2p$-block) considered [this is 
in an extended version 
[ADC(2)x] \cite{Baldea:2007b}, the results become comparable to the 
equation of motion coupled cluster singles and doubles method (EOM-CCSD) \cite{Nooijen:95a}.
Unfortunately, a code enabling computations 
for molecular sizes of interest for molecular transport is not (yet) available.

The inspection of the tables presented below reveals that the results obtained via 
the state-of-the-art IP- and EA-EOM-CCSD method \cite{Stanton:94,Nooijen:95a}
and aug-cc-pVDZ (Dunning augmented correlation consistent double zeta) sets
can be trusted. This is illustrated both by the good agreement between the EOM-CCSD  
and the $\Delta$-CCSD values and by the fact that corrections due to triples [CCSD(T)] 
yield changes that are irrelevant within numerical errors. For understanding the
impact of polarization and diffuse functions we also present results obtained 
by using other basis sets: cc-pVDZ and cc-pVTZ, Pople basis sets (6-31G*, 6-311G* 6-311++G(d,p)),
Dunning-Huzinaga double- and triple-zeta (DZ (DZ95 in ref.~\cite{g09}), DZP, TZ2P), 
Karlsruhe basis sets (svp, dzp, tzp, qz2p), and 
(merely to compare with earlier MP2-based transport 
calculations, cf.~Table \ref{table-bdt-mp2}) STO-6G.
%%%%%%%%%%%%%%%%%%%%%%%%%%%%%%%%%%%%%%%%%%%%%%%%%%%%%%%%%%%%%%%%%%%%%%%%%%%%%%%% 
\section{Results and discussion}
\label{sec:results}
%%%%%%%%%%%%%%%%%%%%%%%%%%%%%%%%%%%%%%%%%%%%%%%%%%%%%%%%%%%%%%%%%%%%%%%%%%%%%%%%
Because the tables presented in this paper, which contain very detailed information on both the
methods and the basis sets employed, are self-explanatory, below we will only briefly 
emphasize the main aspects related to the 
lowest electron attachment ($EA$) and ionization ($IP$) 
energies of the molecules of interest. Still, as a technical remark, 
let us mention that, in view of the fact that double zeta (DZ) sets 
are ubiquitous in transport studies, among other basis sets, we have always 
included DZ-based results in the relevant tables.
%%%%%%%%%%%%%%%%%%%%%%%%%%%%%%%%%%%%%%%%%%%%%%%%%%%%%%%%%%%%%%%%%%%%%%%%%%%%%%%% 
\subsection{Lowest electron attachment energies}
\label{sec:ea}
%%%%%%%%%%%%%%%%%%%%%%%%%%%%%%%%%%%%%%%%%%%%%%%%%%%%%%%%%%%%%%%%%%%%%%%%%%%%%%%%
Table \ref{table-44bpy}, \ref{table-bdcn}, and \ref{table-2bdcn}
collect results on the lowest electron attachment energies for 
4,4'-bipyridine (\ce{C6H4N}$_2$, 44BPY), 
1,4-dicyanobenzene (\ce{NC}-\ce{C6H4}-\ce{CN}, BDCN), 
and 4,4'-dicyano-1,1'-biphenyl (\ce{NC}-\ce{(C6H4)2}-\ce{CN}, 2BDCN).  

Table \ref{table-44bpy} presents very detailed numerical results for 44BPY, 
a showcase molecule 
\cite{Baldea:2012i,Baldea:2013a,Baldea:2013b,Baldea:2013c,Baldea:2014a},
in order to illustrate the main issues, which we have encountered 
in calculations of electroaffinities for molecules utilized in molecular electronics.
The EA-EOM-CCSD method predicts a weakly bound anion 44BPY$^{\bullet -}$ ($EA \agt 0$). 
The essential condition for this is the inclusion of a sufficient number of
diffuse basis functions. 
As emphasized recently \cite{Baldea:2013a},  
it is not the basis set size that matters: as visible in Table \ref{table-44bpy}, 
the basis set CC-pVTZ, which is larger 
than aug-cc-pVDZ, cannot stabilize the anion, 
just because diffuse functions are missing. 

Notice also that a correct description requires a proper treatment of 
electron correlations. 
Even including sufficient diffuse basis functions (aug-cc-pVDZ), 
the SCF description (both Koopmans theorem and 
$\Delta$-SCF) is even qualitatively inadequate;
the anion is predicted to be unstable. This is 
already known from earlier work 
\cite{Kassab:96,Kassab:98,Baldea:2013a}.
As visible in Table \ref{table-44bpy}, 
we found that electron correlations (which, by definition, measure
deviation from SCF) cannot be adequately included via MP2 and the OVGF 
\cite{Cederbaum:75,Schirmer:84}: the 44BPY$^{\bullet -}$ anion remains unstable.

On the other side, while agreeing among themselves, the CC2, ADC(2), 
and $\Delta$-DFT all overestimate the anion stability. Concerning CC2, one can
remark that $\Delta$-CC2 performs better than EOM-CC2.

Basically, the conclusions formulated above by analyzing the electron affinity 
of 44BPY also holds for the other two molecules [1,4-dicyanobenzene (BDCN) 
and 4,4'-dicyano-1,1'-biphenyl (2BDCN)] investigated; see Table \ref{table-bdcn} and
\ref{table-2bdcn}. Quantitatively, there is an important difference: 
the BDCN$^{\bullet -}$ and 2BDCN$^{\bullet -}$ anions are substantially more stable:
their electroaffinities are by about $0.7$\,eV larger than 
that of the 44BPY$^{\bullet -}$ anion. 
% Concerning the basis sets, one should still note that although qualitatively 
% the anion stability is correctly described
% at aug-cc-pVDZ level, larger basis sets are desirable for a quantitative 
% description: at aug-cc-pVTZ level, the EAs become by almost $0.2$\,eV larger.

To summarize, the anions considered above can be accurately described at the
CCSD level of theory provided that the basis set employed includes sufficient 
diffuse functions: 
triplet corrections [\emph{i.e.}, CCSD(T)] merely yield 
modifications of the EA-values within numerical errors, and the 
EA-EOM-CCSD values agree well with the $\Delta$-CCSD values. 
%%%%%%%%%%%%%%%%%%%%%%%%%%%%%%%%%%%%%%%%%%%%%%%%%%%%%%%%%%%%%%%%%%%%%%%%%%%%%%%% 
\subsection{Lowest ionization energies}
\label{sec:ip}
%%%%%%%%%%%%%%%%%%%%%%%%%%%%%%%%%%%%%%%%%%%%%%%%%%%%%%%%%%%%%%%%%%%%%%%%%%%%%%%%
As representatives of molecules embedded in nanojunctions exhibiting a 
HOMO-mediated (p-type) conduction, we have studied and present detailed 
results for benzenedithiol (\ce{HS}-\ce{C6H4}-\ce{SH}, BDT, 
Table \ref{table-bdt}) and related molecules 
(\ce{S}-\ce{C6H4}-\ce{S} and \ce{S}-\ce{C6H3F}-\ce{S},
Table \ref{table-bdt-mp2}), 
dibenzenedithiol (\ce{HS}-\ce{(C6H4)2}-\ce{SH}, 2BDT, Table \ref{table-2bdt}), 
1,6-hexanemonothiol 
(\ce{H}-\ce{(CH2)6}-\ce{SH}, C6MT, Table \ref{table-c6t}), and 1,6-hexanedithiol 
(\ce{HS}-\ce{(CH2)6}-\ce{SH}, C6DT, Table \ref{table-c6dt}). 

As visible in these tables, 
the lowest ionization energies can be estimated with a good relative accuracy 
($\alt 4\%$), which is satisfactory for quantum chemical calculations for many
purposes, within IP-EOM-CCSD calculations by using rather modest basis sets.  
Although not dramatically large, the corresponding absolute error 
($\alt 0.4$\,eV) is still non-negligible from a molecular transport perspective
in view of the rather small energy offset of the dominant molecular orbital relative 
to electrodes' Fermi level. So, good basis sets are required not only for $EA$'s, 
but also for an adequate $IP$'s. Even for cations, the various lower level 
many-body approximations 
(MP2, CC2, ADC(2), CCSD(2) as well as their $\Delta$-versions), deviating by 
up to $\sim 0.4$\,eV from the EOM-CCSD approach, are still not too satisfactory. 

In view of the present results, the $\Delta$-DFT method cannot be recommended: 
deviations from the $IP$-EOM-CCSD estimate can be very large;
the example presented in Table \ref{table-c6dt} indicates an error of $\sim 0.8$\,eV.

A special mention deserves the OVGF approximation, 
which appears to provide the $IP$-estimates closest to EOM-CCSD; 
the differences are smaller than $0.1$\,eV.
%%%%%%%%%%%%%%%%%%%%%%%%%%%%%%%%%%%%%%%%%%%%%%%%%%%%%%%%%%%%%%%%%%%%%%%%%%%%%%%
\subsection{Comparison with results of previously utilized many-body methods}
\label{sec:basis}
%%%%%%%%%%%%%%%%%%%%%%%%%%%%%%%%%%%%%%%%%%%%%%%%%%%%%%%%%%%%%%%%%%%%%%%%%%%%%%%
The foregoing analysis drew attention that both the quantum chemical 
method and the basis set utilized are important for correct 
$EA$- and $IP$-estimations. Employing small basis sets 
is particularly tempting for truly \emph{ab initio} approaches, which are otherwise
impracticable, as they require much more
RAM-memory, disc space, and computational time than ubiquitous DFT-flavors. 

In this subsection, we will scrutinize the reliability  
of the results obtained by post-DFT approaches reported in two earlier
studies \cite{Shimazaki:09,Thygesen:11c}. To assess the validity of those methods,
we will present a comparison with results of the present methods also using 
the same small basis sets of ref.~ \cite{Shimazaki:09,Thygesen:11c}.

Ref.~\cite{Shimazaki:09} reported results obtained within an MP2-like approach 
for two molecules, namely (\ce{S}-\ce{C6H4}-\ce{S} and \ce{S}-\ce{C6H3F}-\ce{S}), 
which are similar to BDT(=\ce{HS}-\ce{C6H4}-\ce{SH}). 
In Table \ref{table-bdt-mp2}, we present results for these molecules
obtained within the methods described above along with those extracted from 
ref.~\cite{Shimazaki:09}. Following ref.~\cite{Shimazaki:09}, we refer
to the latter results as ``MP2-based''. 
Still, for clarity, we should note that, 
in the present terminology, the method of ref.~\cite{Shimazaki:09}
coincides with that labeled ``2nd-order pole'' here, 
as it corresponds to the second-order correction in the the electron-electron 
interaction to the electronic self-energy \cite{Cederbaum:75,Schirmer:84}.
So, the results labeled ``MP2-based'' and ``2nd-order pole'' in Table \ref{table-bdt-mp2} 
should coincide. They should but they do not coincide; or, more precisely,
they are substantially different. We cannot understand these large differences ($\agt 1$\,eV)
visible in Table \ref{table-bdt-mp2}.
The only thing not specified in ref.~\cite{Shimazaki:09} is the molecular geometry
utilized in the calculations. However, as actually expected, 
the results presented in Table \ref{table-bdt-mp2} reveal that 
differences in (optimized) geometries have a considerable smaller 
impact (maybe $\sim 0.2$\,eV). Most importantly, as seen in Table \ref{table-bdt-mp2}, 
it is the combined effect of an inaccurate method
and a too small basis set that results in very large errors ($\sim 3$\,eV) 
for $IP$'s. 

By using the same (DZ) basis sets as in ref.~\cite{Thygesen:11c}, 
Table \ref{table-bdt-gw} demonstrates that the results deduced within 
the $GW$ method are not adequate to estimate the energies of the frontier orbitals
with the accuracy required for molecular transport studies. 
Differences $\sim 0.5$\,eV 
between the $GW$ and EOM-CCSD visible in Table \ref{table-bdt-gw} are too large,
given the fact that for this molecule (BDT) the HOMO energy offset directly measured by 
ultraviolet photoelectron spectroscopy (UPS)
amounts to $\simeq 1$\,eV \cite{Frisbie:11}. Drawbacks of $GW$-based transport 
approaches were previously pointed out \cite{Millis:09}.
%%%%%%%%%%%%%%%%%%%%%%%%%%%%%%%%%%%%%%%%%%%%%%%%%%%%%%%%%%%%%%%%%%%%%%%%%%%%%%%% 
\subsection{Spatial distribution of the frontier molecular orbitals}
\label{sec:spatial-extension}
%%%%%%%%%%%%%%%%%%%%%%%%%%%%%%%%%%%%%%%%%%%%%%%%%%%%%%%%%%%%%%%%%%%%%%%%%%%%%%%%
As highlighted above, both the method and the basis sets 
employed are essential to properly estimate the energy of the frontier orbitals of an isolated
molecule. Still, however important, the level energy of an isolated molecule is 
not the whole issue. In a molecular junction, the active molecule is linked to electrodes,
which yield shifts in energy via image charge effects. This effect, which is well established 
in surface science \cite{desjonqueres:96}, was embodied in recent studies on molecular electronics 
in a simplified form, namely by assuming point-like molecular orbitals midway between
electrodes \cite{Neaton:06,Venkataraman:12a}. This assumption is indeed a comfortable  
approximation, as it can readily be implemented in one-shot DFT+$\Sigma$
transport calculations \cite{Louie:07,Venkataraman:12a}. 

The obvious critical point here, on which ref.~\cite{Baldea:2013b} has recently 
drawn attention, is to what extent is it legitimate to approximate a real molecular orbital 
as a point charge. Within classical electrostatics, the interaction energy of an electron located at $z$ 
with the image charges created in two infinite planar electrodes can be 
exactly expressed by \cite{Sommerfeld:33,Baldea:2012f}
%%%%%%%%%%%%%%%%%%%%%%%%%%%%%%%%%%%%%%%%%%%%%%%%%%%%%%%%%%%%%%%%%%%%%%%%%%%%%%
\begin{equation}
\label{eq-phi-i-0}
\phi_{i}(z) =
\frac{e^2}{4 d}
\left[
- 2 \psi(1)  + \psi\left(\frac{z - z_s}{d}\right) + \psi\left(\frac{z_t - z}{d}\right)
\right].
\end{equation}
%%%%%%%%%%%%%%%%%%%%%%%%%%%%%%%%%%%%%%%%%%%%%%%%%%%%%%%%%%%%%%%%%%%%%%%%%%%%%%
Here, $z_{s,t}$ are the positions of the image planes (which are slightly shifted 
from the real electrodes, see, \emph{e.g}, ref.~\cite{Baldea:2013b} and citations therein), 
$d \equiv z_t - z_s$, and $\psi(z) \equiv d\, \log \Gamma(z) / d\, z$ is the
digamma function. For a real molecular orbital, the image-driven energy shift
should be computed by weighting eqn~(\ref{eq-phi-i-0}) with the MO-spatial density
$\rho_{MO} = \vert \Psi_{MO}\vert^2$, which is determined by its wave function $ \Psi_{MO}$
%%%%%%%%%%%%%%%%%%%%%%%%%%%%%%%%%%%%%%%%%%%%%%%%%%%%%%%%%%%%%%%%%%%%%%%%%%%%%%
\begin{equation}
\label{eq-vIm}
\Phi = \int_{z_s}^{z_t} d\,z \,  \rho_{1D}^{MO}(z)\, \phi_{i} (z) ,
\end{equation}
%%%%%%%%%%%%%%%%%%%%%%%%%%%%%%%%%%%%%%%%%%%%%%%%%%%%%%%%%%%%%%%%%%%%%%%%%%%%%%
%%%%%%%%%%%%%%%%%%%%%%%%%%%%%%%%%%%%%%%%%%%%%%%%%%%%%%%%%%%%%%%%%%%%%%%%%%%%%%
\begin{equation}
\label{eq-rho1}
\rho_{1D}^{MO} (z) =  \int \hspace*{-1.2ex} \int d\,x\,d\,y \,
\vert \Psi_{MO}(x, y, z)\vert^2 .
\end{equation} 
%%%%%%%%%%%%%%%%%%%%%%%%%%%%%%%%%%%%%%%%%%%%%%%%%%%%%%%%%%%%%%%%%%%%%%%%%%%%%%
Close to electrodes, e.~g., $z \agt z_s$, eqn~(\ref{eq-phi-i-0}) recovers the
classical expression
%%%%%%%%%%%%%%%%%%%%%%%%%%%%%%%%%%%%%%%%%%%%%%%%%%%%%%%%%%%%%%%%%%%%%%%%%%%%%%
\begin{equation}
\label{eq-phi-i-1}
% {\raisebox{-0.3ex}{$\stackrel{<}{\sim}$}}
{\phi}_{i}(z) {\raisebox{-0.3ex}{$\stackrel{ z \raisebox{-0.3ex} \agt z_s}{\approx}$}}
\phi_{i, cl} (z) = - \frac{e^2}{4 (z - z_s)}
\end{equation}
%%%%%%%%%%%%%%%%%%%%%%%%%%%%%%%%%%%%%%%%%%%%%%%%%%%%%%%%%%%%%%%%%%%%%%%%%%%%%%
for a single image plane. This demonstrates that, in cases of molecular orbitals
with significant spatial extension, the main contributions to the image-driven 
energy shift $\Phi$ come from regions close to electrodes.

Since, except for ref.~\cite{Baldea:2013b}, 
the spatial extension of the dominant molecular orbitals did not received consideration
in previous studies, we have decided to 
systematically investigate this aspect for typical molecules of interest for 
nanotransport. 
To this aim, inspecting spatial densities of 
(completely unphysical LUMO) Kohn-Sham orbitals 
makes little sense, and (especially LUMO) HF orbitals may represent a 
too crude approximation. Therefore, like in ref.~\cite{Klaiman:13}, 
we have calculated the natural orbital expansion of the 
corresponding reduced density matrices at the EOM-CCSD level.
 
This is the most reliable approach to characterize
the spatial distribution of the extra electron 
or hole in molecules with
n-type (LUMO-mediated) or p-type (HOMO-mediated) conduction, and we are not aware
of a similar study conducted in conjunction with molecular transport 
at this level of theory. For all the molecules considered,
by inspecting the natural orbital expansion, we found that 
the extra electron or hole is almost entirely 
($> 97\%$) concentrated in a single natural orbital. 

Most importantly from the present standpoint, we found not even a
molecule whose dominant MO reside in a very narrow spatial region around the
center.

Rather than being strongly localized close to the center, 
in cases of n-conduction,
we found that the natural orbital of the extra electron  
is more or less uniformly spread over the whole molecule. This is illustrated by 
the examples depicted in
Fig.~\ref{fig:lumo-44bpy}
% , \ref{fig:lumo-bdcn}, 
and \ref{fig:lumo-2bdcn}. 
%%%%%%%%%%%%%%%%%%%%%%%%%%%%%%%%%%%%%%%%%%%%%%%%%%%%%%%%%%%%%%%%%%%%%%%%%%%%%%%% 
\begin{figure}[h!]
$ $\\[0ex]
% \centerline{\hspace*{0.ex}\includegraphics[width=0.5\textwidth,angle=0]{44bpy_LUMO_cfour_augccpvdz.eps}\hspace*{0.ex}}
% \centerline{\hspace*{0.ex}\includegraphics[width=0.5\textwidth,angle=0]{Fig1.eps}\hspace*{0.ex}}
% \centerline{\hspace*{0.ex}\includegraphics[width=0.5\textwidth,angle=0]{44bpy_LUMO_gimp.eps}\hspace*{0.ex}}
\centerline{\hspace*{0.ex}\includegraphics[width=0.45\textwidth,angle=0]{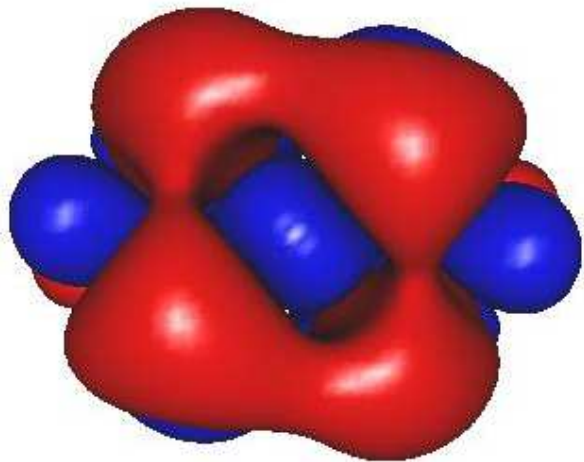}\hspace*{0.ex}}
$ $\\[-1.4ex]
\caption{The almost singly occupied natural orbital corresponding to the anion's extra electron 
of the 44BPY$^{\bullet -}$ anion (``LUMO'')
obtained via EA-EOM-CCSD/aug-cc-pVDZ calculations is delocalized over the whole molecule.}
\label{fig:lumo-44bpy}
\end{figure}
%%%%%%%%%%%%%%%%%%%%%%%%%%%%%%%%%%%%%%%%%%%%%%%%%%%%%%%%%%%%%%%%%%%%%%%%%%%%%%%%
%%%%%%%%%%%%%%%%%%%%%%%%%%%%%%%%%%%%%%%%%%%%%%%%%%%%%%%%%%%%%%%%%%%%%%%%%%%%%%%% 
\begin{figure}[h!]
$ $\\[6ex]
% \centerline{\hspace*{0.ex}\includegraphics[width=0.4\textwidth,angle=0]{2bdcn_LUMO_cfour_augccpvdz.eps}\hspace*{0.ex}}
% \centerline{\hspace*{0.ex}\includegraphics[width=0.5\textwidth,angle=0]{Fig2.eps}\hspace*{0.ex}}
% \centerline{\hspace*{0.ex}\includegraphics[width=0.4\textwidth,angle=0]{2bdcn_LUMO_gimp.eps}\hspace*{0.ex}}
\centerline{\hspace*{0.ex}\includegraphics[width=0.45\textwidth,angle=0]{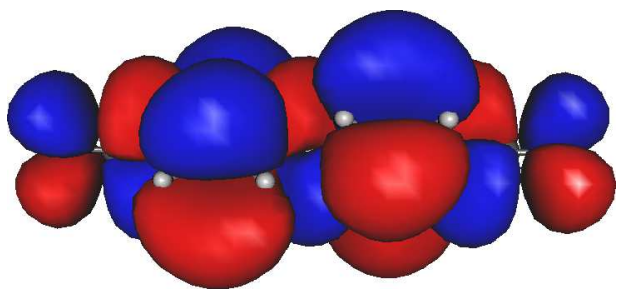}\hspace*{0.ex}}
$ $\\[-1.4ex]
\caption{The almost singly occupied natural orbital of the extra electron in the 
2BDCN anion (``LUMO'') is delocalized
over the whole molecule. Result of EA-EOM-CCSD/aug-cc-pVDZ calculations.}
\label{fig:lumo-2bdcn}
\end{figure}
%%%%%%%%%%%%%%%%%%%%%%%%%%%%%%%%%%%%%%%%%%%%%%%%%%%%%%%%%%%%%%%%%%%%%%%%%%%%%%%%
(We employed Gabedit \cite{gabedit} to generate the figures presented in this paper.)
Therefore, the difference between the value $\Phi_{loc} = - e^2 \ln 2/d$
obtained by setting $z = (z_s + z_t)/2$ in eqn~(\ref{eq-phi-i-0}), which
corresponds to a point-like MO located in the middle of two infinite metallic plates
\cite{Venkataraman:12a}, 
is substantially different from the value $\Phi_r$ deduced via eqn~(\ref{eq-vIm}) 
by using the realistic natural orbital density. For the
44BPY molecule, the values thus obtained for the image-driven LUMO 
shifts are $\Phi_r = -2.13$\,eV and 
$\Phi_{loc} = -1.16$\,eV. A difference of about 1\,eV is a big effect.

We have also computed spatial distributions of the natural orbital 
of the extra hole (``HOMO'') in cation species relevant for molecules 
exhibiting p-type conduction. The 
examples presented in 
% Fig.~\ref{fig:homo-bdt}, 
Fig.~\ref{fig:homo-2bdt} and 
\ref{fig:homo-c6t}
% , and \ref{fig:homo-c6dt} 
illustrate two different behaviors, 
which we found to be characteristic for HOMO distributions. 
%%%%%%%%%%%%%%%%%%%%%%%%%%%%%%%%%%%%%%%%%%%%%%%%%%%%%%%%%%%%%%%%%%%%%%%%%%%%%%%% 
\begin{figure}[h!]
$ $\\[6ex]
% \centerline{\hspace*{0.ex}\includegraphics[width=0.4\textwidth,angle=0]{2bdt_HOMO_cfour_augccpvdz.eps}\hspace*{0.ex}}
% \centerline{\hspace*{0.ex}\includegraphics[width=0.5\textwidth,angle=0]{Fig3.eps}\hspace*{0.ex}}
% \centerline{\hspace*{0.ex}\includegraphics[width=0.4\textwidth,angle=0]{2bdt_HOMO_gimp.eps}\hspace*{0.ex}}
\centerline{\hspace*{0.ex}\includegraphics[width=0.45\textwidth,angle=0]{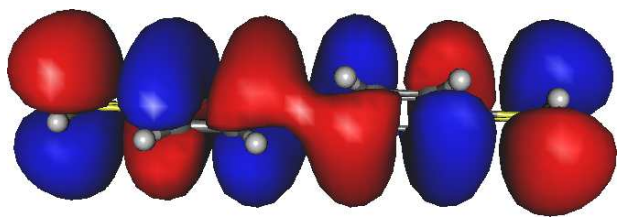}\hspace*{0.ex}}
$ $\\[-1.4ex]
\caption{The almost singly occupied natural orbital of the extra hole 
(``HOMO'') 
in the BDT cation 
is delocalized over the whole molecule. Result of IP-EOM-CCSD/aug-cc-pVDZ calculations.}
\label{fig:homo-2bdt}
\end{figure}
%%%%%%%%%%%%%%%%%%%%%%%%%%%%%%%%%%%%%%%%%%%%%%%%%%%%%%%%%%%%%%%%%%%%%%%%%%%%%%%%
%%%%%%%%%%%%%%%%%%%%%%%%%%%%%%%%%%%%%%%%%%%%%%%%%%%%%%%%%%%%%%%%%%%%%%%%%%%%%%%% 
\begin{figure}[h!]
$ $\\[6ex]
% \centerline{\hspace*{0.ex}\includegraphics[width=0.4\textwidth,angle=0]{HOMO_c6t_cfour_augccpvdz.eps}\hspace*{0.ex}}
% \centerline{\hspace*{0.ex}\includegraphics[width=0.5\textwidth,angle=0]{Fig4.eps}\hspace*{0.ex}}
% \centerline{\hspace*{0.ex}\includegraphics[width=0.4\textwidth,angle=0]{HOMO_c6t_gimp.eps}\hspace*{0.ex}}
\centerline{\hspace*{0.ex}\includegraphics[width=0.45\textwidth,angle=0]{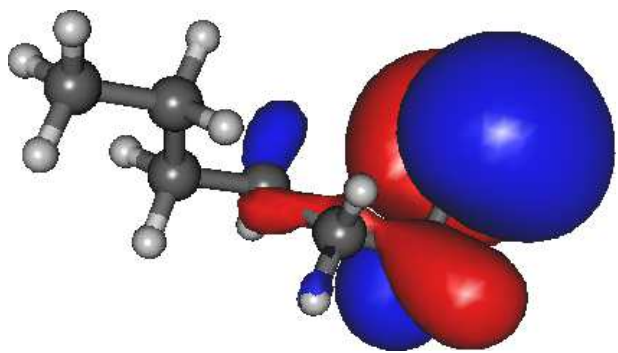}\hspace*{0.ex}}
$ $\\[-1.4ex]
\caption{The almost singly occupied natural orbital of the extra hole 
(``HOMO'') 
in the 
hexanemonothiol cation is localized in the vicinity of the sulfur atom. 
Result of IP-EOM-CCSD/aug-cc-pVDZ calculations.}
\label{fig:homo-c6t}
\end{figure}
%%%%%%%%%%%%%%%%%%%%%%%%%%%%%%%%%%%%%%%%%%%%%%%%%%%%%%%%%%%%%%%%%%%%%%%%%%%%%%%%
On one side, we found HOMO's like that presented in 
% Fig.~\ref{fig:homo-bdt} and
Fig.~\ref{fig:homo-2bdt}, which are similar to the LUMO's discussed above; 
they are delocalized over the whole molecule. On the other side, we found
HOMO's strongly localized on the anchoring groups. Typical for this behavior 
are alkanemono- and dithiols, as illustrated in Fig.~\ref{fig:homo-c6t} 
% and \ref{fig:homo-c6dt}.

To obtain the above value $\Phi_r$($=-2.16$\,eV) for 44BPY, we applied 
the cutoff procedure near electrodes described in detail in ref.~\cite{Baldea:2013b}.
Results of preliminary calculations with metal atoms linked to 
the molecules with LUMO-mediated conduction considered here indicate that, like the case of 
ref.~\cite{Baldea:2013b}, the spatial density LUMO 
(single occupied natural orbital corresponding to the
extra electron) does not substantially penetrate into electrodes. This is important: 
corrections due to image charges are not
dramatically affected by the cutoff procedure close to electrodes.
This behavior contrasts to that of the HOMO's. Whether delocalized 
(like that of 
% Fig.~\ref{fig:homo-bdt})
Fig.~\ref{fig:homo-2bdt}) 
or localized (like that from Fig.~\ref{fig:homo-c6t}) 
% and \ref{fig:homo-c6dt}) 
on the terminal groups of the
\emph{isolated} molecules, we found that 
the HOMO distributions substantially penetrates into
electrodes. Although we only checked that this happens for molecules of the types 
discussed above, we believe that this is a general HOMO property that ensures stable
molecule-electrode bonds. A cutoff procedure is needed to eliminate spurious 
divergences of the classical expression of the interaction energy with image charges
at $z=z_{s,t}$ [\emph{cf.~}eqn (\ref{eq-phi-i-0}) and (\ref{eq-phi-i-1})], which 
ignores quantum mechanical effects and electrodes' atomistic structure. 
A possible cutoff procedure consists of multiplying
the classical expression with factors $1 - \exp(-\mu\vert z -z_{s,t})$
({$\propto \vert z - z_ {s,t} \vert$ as $z \to z_{s,t}$); 
see, \emph{e.g}, ref.~\cite{Baldea:2013b} and citations therein. Cutoff procedures 
make sense only 
if the results do not sensitively depend on the value of cutoff parameter $\mu$,
and this cannot be the case if (HO)MO densities have non-negligible values 
in spatial regions close to electrodes ($z\sim z_{s,t}$).
%%%%%%%%%%%%%%%%%%%%%%%%%%%%%%%%%%%%%%%%%%%%%%%%%%%%%%%%%%%%%%%%%%%%%%%%%%%%%%%% 
\subsection{The exciton binding energy as evidence for important electron correlations}
\label{sec:be}
%%%%%%%%%%%%%%%%%%%%%%%%%%%%%%%%%%%%%%%%%%%%%%%%%%%%%%%%%%%%%%%%%%%%%%%%%%%%%%%%
As a possible way to quantify electron correlations, the solid-state community 
employs the difference between the so-called charge gap $\Delta_c$ and the optical gap(=lowest
excitation energy) $\Delta_o$; 
see \emph{e.g.}~ref.~\cite{Baldea:2010a,Baldea:2012i} and citations therein.
The charge gap, which is what molecular physicists normally call the HOMO-LUMO gap,
can be expressed as $\Delta_c = IP-EA$. 
Loosely speaking [because in reality the single-particle (MO) picture breaks down 
in the exciton problem], the difference between the charge gap and the 
optical gap is that, in determining $EA$, both 
HOMO and LUMO are occupied. By contrast, in the lowest optical excitation, the HOMO becomes
empty as the LUMO becomes occupied. $\Delta_c$ should be larger than $\Delta_o$
because of the (negative) attraction energy between the oppositely charged electron (LUMO) and hole (HOMO)
This difference is
referred to as the exciton binding energy $EBE$ (see, \emph{e.g.}, ref.~\cite{mahan}) 
%%%%%%%%%%%%%%%%%%%%%%%%%%%%%%%%%%%%%%%%%%%%%%%%%%%%%%%%%%%%%%%%%%%%%%%%%%%%%%%%
\begin{equation}
\label{eq-be}
EBE = \Delta_c - \Delta_o .
\end{equation}
%%%%%%%%%%%%%%%%%%%%%%%%%%%%%%%%%%%%%%%%%%%%%%%%%%%%%%%%%%%%%%%%%%%%%%%%%%%%%%%%
The various tables of the present paper 
only include results for $EBE$ obtained within the most 
accurate method utilized (EOM-CCSD).
The $EBE$-values shown there are substantial, 
amounting to up to $\sim 50\% $ of the charge gap. As expected for molecules with aromatic units
and delocalized electrons, the $EBE$ decreases with increasing molecular size; 
compare the $EBE$-values for BDCN (Table \ref{table-bdcn}) and 2BDCN (Table \ref{table-2bdcn}),
and for BDT (Table \ref{table-bdt}) and 2BDT (Table \ref{table-2bdt}). The 
$EBE$-values ($EBE \sim 3.5 - 4.6$\,eV) estimated for all the molecules analyzed in the present paper 
are substantially larger than, \emph{e.g.}, for $\pi$-conjugated organic thin films 
($EBE \sim 0.6 - 1.4$\,eV) \cite{Zangmeister:04}. 

So, one should conclude that, for species of interest for molecular electronics, 
electron correlations are very strong. This aspect may be quite relevant for developing 
correlated transport approaches \cite{Baldea:2010f}; \emph{e.g.}, even if the charge transport is dominated by
the LUMO, an electron traveling through the molecule can interact
with the HOMO.  
%%%%%%%%%%%%%%%%%%%%%%%%%%%%%%%%%%%%%%%%%%%%%%%%%%%%%%%%%%%%%%%%%%%%%%%%%%%%%%%% 
\section{Conclusion}
\label{sec:conclusion}
%%%%%%%%%%%%%%%%%%%%%%%%%%%%%%%%%%%%%%%%%%%%%%%%%%%%%%%%%%%%%%%%%%%%%%%%%%%%%%%%
In this paper, we have presented benchmark quantum chemical 
calculations for the lowest electron attachment
and ionization energies of several isolated molecular species of interest
for molecular electronics. 
In assessing the importance of the accuracy to be achieved by estimating $EA$'s and $IP$'s, 
it is worth mentioning that a proper understanding of the charge transport at nanoscale
does not only mean to reproduce the (order of) magnitude of the currents (which can be adjusted by
``manipulating'' both $\varepsilon_0$ and the broadening functions $\Gamma$ not considered here), but also 
the biases $V$ characterizing a non-Ohmic regime, which is basically determined by $\varepsilon_0$  
($eV ~ \varepsilon_0$, see \cite{Baldea:2012b}).

The main results presented above can be summarized as follows:

(i) For all molecules, the differences between the HF-MO energy 
(Koopmans theorem) and $\Delta$-SCF values are 
large. This demonstrates that orbital relaxation is substantial. Electron correlations
are also important, as revealed by important departures from the SCF results as well as 
by substantial differences between the various post-SCF methods considered.

(ii) The present results demonstrate the need both for accurate methods and good basis sets
beyond those currently utilized in transport approaches. In particular, 
employing basis sets with sufficient diffuse functions is essential to correctly describe
electron affinities.

(iii) Kohn-Sham orbital energies can by no means be used to estimate $IP$'s and $EA$'s. Even with 
$\Delta$-corrections, DFT-based methods do not appear to achieve the desired accuracy
of estimating the relevant MO energy offsets $\varepsilon_0$. As visible in the various tables,  
because $\Delta$-DFT estimates are much weakly dependent on the basis sets that those of EOM-CCSD; so, 
$\Delta$-DFT may convey a false impression on the importance of the basis sets to be utilized in calculations.

(iv) MP2-based methods appear to be completely inadequate for describing anions. 
For cations, they may yield substantially different results; 
\emph{e.g.}, compare the deduced via 
$\Delta$-MP2 and the second-order order correction to self-energy (``2nd-order pole) 
($>1.6$\,eV in Table~\ref{table-2bdt}). Examples showing that different methods to 
include second-order terms in the electron-electron interaction in other contexts 
were discussed earlier \cite{Holleboom:90}.  

(v) For the presently investigated molecules exhibiting p-type conduction, 
the OVGF method  
represents an excellent compromise in terms of 
computational effort and accuracy of $IP$-estimates. Unlike the other diagrammatic methods 
considered here ($GW$ and $ADC$), the OVGF method 
does not require to self-consistently solve a(n integral) Dyson equation; the electron self-energy
$\boldsymbol{\Sigma}(\varepsilon)$ 
can be expressed in closed analytical form, and what needs to solve is a nonlinear algebraic equation 
for $\varepsilon$ \cite{Schirmer:84}. To be fair, let us also mention that, for the 
presently considered molecules that form junctions characterized by
n-type conduction, the OVGF method turned out to be totally inadequate.

(vi) The spatial distribution of the frontier orbitals plays an important role to reliably estimate 
image-driven shifts of the relevant MO-energies.
To the best of our knowledge, this is the first systematic study on the spatial distribution 
of the extra electron or hole in molecular species of interest for molecular transport 
at this (EOM-CCSD natural orbital expansion) level of theory. None of the molecules considered in this paper 
was found to possess point-like frontier molecular orbitals, a fact that contradicts 
the common assumption made in the field.
%%%%%%%%%%%%%%%%%%%%%%%%%%%%%%%%%%%%%%%%%%%%%%%%%%%%%%%%%%%%%%%%%%%%%%%%%%%%%%%%
%%%%%%%%%%%%%%%%%%%%%%%%%%%%%%%%%%%%%%%%%%%%%%%%%%%%%%%%%%%%%%%%%%%%%%%%%%%%%%%% 
%%%%%%%%%%%%%%%%%%%%%%%%%%%%%%%%%%%%%%%%%%%%%%%%%%%%%%%%%%%%%%%%%%%%%%%%%%%%%%%%
\subsubsection*{Acknowledgments~~}
The author thanks 
Shachar Klaiman and Evgeniy Gromov for invaluable help to perform quantum chemical calculations,
and Jochen Schirmer for useful discussions.
Thanks are also due to Vitja Vysotskiy, 
whose fully parallelized PRICD-$\Sigma$(2) code 
has been employed to obtain the ADC(2) results for 
electron affinity and ionization energies reported here.
Calculations for this work have been partially done on the 
high performance bwGrid cluster \cite{bwGrid}.
Financial support from the Deu\-tsche For\-schungs\-ge\-mein\-schaft 
(grant BA 1799/2-1) is gratefully acknowledged.\\
% {\bf Electronic Supplementary Information {\esi}:} Supplementary material (computational details and tables) is available in the online version of this article at http://\ldots.
%%%%%%%%%%%%%%%%%%%%%%%%%%%%%%%%%%%%%%%%%%%%%%%%%%%%%%%%%%%%%%%%%%%%%%%%%%%%%%%%
\renewcommand\refname{Notes and references}
\footnotesize{
\bibliographystyle{rsc}
% \bibliography{/home/ioan/QDs/LinearResponse/paper/bibl,/home/ioan/QDs/LinearResponse/paper/misc,/home/ioan/QDs/LinearResponse/paper/arxiv}
\bibliography{manuscript} 

\begin{mcitethebibliography}{64}
\bibitem[dys()]{dyson}
Notice that, via the Dyson equation, each term included in the self-energy
  amounts to include an infinite sub-series of terms in the perturbation
  expansion.\relax
\mciteBstWouldAddEndPunctfalse
\mciteSetBstMidEndSepPunct{\mcitedefaultmidpunct}
{}{\mcitedefaultseppunct}\relax
\EndOfBibitem
\bibitem[why()]{why-not-aug-cc-pvdz}
We used the larger 6-311++G(d,p) basis set for this molecule instead of the
  aug-cc-pVDZ basis set employed in all other cases merely because it created
  convergence problems.\relax
\mciteBstWouldAddEndPunctfalse
\mciteSetBstMidEndSepPunct{\mcitedefaultmidpunct}
{}{\mcitedefaultseppunct}\relax
\EndOfBibitem
\end{mcitethebibliography}
% 
% \end{document}
%%%%%%%%%%%%%%%%%%%%%%%%%%%%%%%%%%%%%%%%%%%%%%%%%%%%%%%%%%%%%%%%%%%%%%%%%%%%%%%
\normalsize
%%%%%%%%%%%%%%%%%%%%%%%%%%%%%%%%%%%%%%%%%%%%%%%%%%%%%%%%%%%%%%%%%%%%%%%%%%%%%%%
%%%%%%%%%%%%%%%%%%%%%%%%%%%%%%%%%%%%%%%%%%%%%%%%%%%%%%%%%%%%%%%%%%%%%%%%%%%%%%%% 
%%%%%%%%%%%%%%%%%%%%%%%%%%%%%%%%%%%%%%%%%%%%%%%%%%%%%%%%%%%%%%%%%%%%%%%%%%%%%%%%
% \end{document}
%%%%%%%%%%%%%%%%%%%%%%%%%%%%%%%%%%%%%%%%%%%%%%%%%%%%%%%%%%%%%%%%%%%%%%%%%%%%%%%%%%%%%%%%%
\begin{table*} % [h!]
\small
\begin{center}
\begin{tabular*}{0.58\textwidth}{@{\extracolsep{\fill}}lllll}
\hline
44BPY/Method   & Basis set   & No.\ functions & $EA$\,(eV) & $EBE$\,(eV)\\
\hline
EOM-CCSD        & DZ          & 136 & -0.684   & \\
EOM-CCSD        & 6-31G**     & 208 & -0.686   & \\
EOM-CCSD        & svp         & 208 & -0.421   & \\
EOM-CCSD        & cc-pVDZ     & 208 & -0.484   & \\
EOM-CCSD        & dzp         & 220 & -0.353   & \\
EOM-CCSD        & 6-311G**    & 264 & -0.340   & \\
EOM-CCSD        & tzp         & 276 & -0.290   & \\
EOM-CCSD        & TZ2P        & 360 & -0.0247  & \\ 
EOM-CCSD        & qz2p        & 416 & -0.0123  & \\
EOM-CCSD        & cc-pVTZ     & 472 & -0.438   & \\
EOM-CCSD        & aug-cc-pVDZ & 348 & 0.0322   & 4.487 \\
EOM-CCSD(T)     & aug-cc-pVDZ & 348 & 0.0293   & \\
                &             &     &          & \\
EOM-CC2         & aug-cc-pVDZ & 348 & 0.360    & \\
ADC(2)          & aug-cc-pVDZ & 348 & 0.370    & \\
                &             &     &          & \\
$\Delta$-SCF & aug-cc-pVDZ & 348 & -0.564 & \\
$\Delta$-MP2       & aug-cc-pVDZ & 348 & -0.500 &  \\
$\Delta$-CCSD      & aug-cc-pVDZ & 348 & 0.0043 &  \\
% \ib{$\Delta$-CC2}       & aug-cc-pVDZ & 348 &        &  \\
                &             &     &          \\
$\Delta$-DFT/B3LYP & DZ    & 136 & 0.185 &   \\
$\Delta$-DFT & 6-31G*      & 196 & -0.126 &    \\
$\Delta$-DFT/B3LYP & svp   & 208 & 0.124 &   \\
$\Delta$-DFT & 6-31G**     & 208 & -0.115 &    \\
$\Delta$-DFT & cc-pVDZ     & 208 & 0.0290 &   \\
$\Delta$-DFT & 6-31+G(d)   & 244 & 0.396 &    \\
$\Delta$-DFT & 6-311G**    & 264 & 0.188 &   \\
$\Delta$-DFT & 6-31+G(d,p) & 268 & 0.407 &    \\
$\Delta$-DFT & 6-31++G(d,p)  & 276 & 0.407 &  \\
$\Delta$-DFT/B3LYP & cc-pVTZ & 472 & 0.281 &   \\
$\Delta$-DFT & aug-cc-pVDZ & 348 & 0.444 &   \\
$\Delta$-DFT & aug-cc-pVTZ & 736 & 0.467 &   \\
                &             &     &          & \\
-KS-LUMO & aug-cc-pVDZ       & 348 &  2.068 & \\    
Koopmans theorem & aug-cc-pVDZ       & 348 & -0.832 & \\    
2nd-order pole (98.3\%)  & aug-cc-pVDZ & 348 & -0.466 & \\
3rd-order pole (98.6\%)  & aug-cc-pVDZ & 348 & -0.562 & \\
OVGF (98.5\% )           & aug-cc-pVDZ & 348 & -0.540 & \\
\hline
\end{tabular*}
\caption{The lowest electron affinity ($EA$) of 4,4$^\prime$-bipyridine (44BPY)
computed by various methods and the exciton binding energy ($EBE$). 
The geometry of the neutral molecule has been 
optimized at DFT/B3LYP/aug-cc-pVDZ level. 
}
\label{table-44bpy}
\end{center}
\end{table*}
%%%%%%%%%%%%%%%%%%%%%%%%%%%%%%%%%%%%%%%%%%%%%%%%%%%%%%%%%%%%%%%%%%%%%%%%%%%%%%%%%%%%%%
%%%%%%%%%%%%%%%%%%%%%%%%%%%%%%%%%%%%%%%%%%%%%%%%%%%%%%%%%%%%%%%%%%%%%%%%%%%%%%%%%%%%%%%%%
\begin{table*} % [h!]
\small
\begin{center}
\begin{tabular*}{0.58\textwidth}{@{\extracolsep{\fill}}lllll}
\hline
BDCN/Method     & Basis set   & No.\ functions & $EA$\,(eV) & $EBE$\,(eV) \\
\hline
EOM-CCSD        & DZ          & 108    & -0.0833 &  \\
EOM-CCSD        & 6-31G**     & 170    & 0.0893   & \\
% EOM-CCSD        & svp         & 160    & 0.330   & \\
% EOM-CCSD        & dzp         & 170    & 0.262   & \\
% EOM-CCSD        & DZP         & 170    & 0.383  & \\
% EOM-CCSD        & cc-pVDZ     & 160    & 0.290  &  \\
% EOM-CCSD        & cc-pVTZ     & 356    & 0.733  & \\
EOM-CCSD        & aug-cc-pVDZ & 266    & 0.717  &  4.614 \\
EOM-CCSD(T)     & aug-cc-pVDZ & 266    & 0.717  &  \\
% EOM-CCSD        & aug-cc-pVTZ & 552    & 0.898  &  \\
                &             &     &          & \\
EOM-CC2         & aug-cc-pVDZ & 266    & 1.047 &   \\
ADC(2)          & aug-cc-pVDZ & 266    & 1.107 &    \\
                &             &     &          & \\
$\Delta$-MP2       & aug-cc-pVDZ & 266 &  0.394 &  \\
$\Delta$-CC2      & aug-cc-pVDZ & 266 & 0.602  &  \\
$\Delta$-CCSD      & aug-cc-pVDZ & 266 & 0.678  &  \\
                &             &     &          & \\
$\Delta$-DFT/B3LYP & aug-cc-pVDZ & 266 & 1.127 &   \\
$\Delta$-DFT/B3LYP & aug-cc-pVTZ & 552 & 1.155 & \\
% $\Delta$-DFT/B3LYP & 6-31G**     & 170 & 0.623  &        \\
% $\Delta$-DFT/B3LYP & 6-311G**    & 204 &  0.924  &  \\
% $\Delta$-DFT/B3LYP & cc-pVDZ     & 160    &  0.742 &        \\
% $\Delta$-DFT/B3LYP & cc-pVTZ     & 356    &  1.018  &       \\
% $\Delta$-DFT/B3LYP & DZ          & 108  &  0.832  &  \\
% $\Delta$-DFT/B3LYP & svp         & 160    &  0.821  &  \\
                &             &     &          & \\
$\Delta$-SCF    & aug-cc-pVDZ & 266 & 0.350    & \\
                &             &     &          & \\
-KS-LUMO  & aug-cc-pVDZ       & 266    &  2.918 & \\
Koopmans theorem & aug-cc-pVDZ       & 266    &  -0.536 & \\    
2nd-order pole (88.6\%)  & aug-cc-pVDZ & 266    & 1.034 & \\
3rd-order pole (91.1\%)  & aug-cc-pVDZ & 266    & 0.253 &   \\
OVGF (90.6\% )           & aug-cc-pVDZ & 266    & 0.445 &      \\
\hline
\end{tabular*}
\caption{
The lowest electron affinity ($EA$) of 1,4-dicyanobenzene (BDCN)
computed by various methods and the exciton binding energy ($EBE$). 
The geometry of the neutral molecule has been 
optimized at DFT/B3LYP/aug-cc-pVDZ level.}
\label{table-bdcn}
\end{center}
\end{table*}
%%%%%%%%%%%%%%%%%%%%%%%%%%%%%%%%%%%%%%%%%%%%%%%%%%%%%%%%%%%%%%%%%%%%%%%%%%%%%%%%%%%%%%
%%%%%%%%%%%%%%%%%%%%%%%%%%%%%%%%%%%%%%%%%%%%%%%%%%%%%%%%%%%%%%%%%%%%%%%%%%%%%%%% 
% \begin{figure}[h!]
% $ $\\[6ex]
% \centerline{\hspace*{0.ex}\includegraphics[width=0.4\textwidth,angle=0]{LUMO_bdcn_cfour_augccpvdz.eps}\hspace*{0.ex}}
% $ $\\[-1.4ex]
% \caption{The almost singly occupied natural orbital of the extra electron in the BDCN
% anion (``LUMO'') is delocalized
% over the whole molecule. Result of EA-EOM-CCSD/aug-cc-pVDZ calculations.}
% \label{fig:lumo-bdcn}
% \end{figure}
%%%%%%%%%%%%%%%%%%%%%%%%%%%%%%%%%%%%%%%%%%%%%%%%%%%%%%%%%%%%%%%%%%%%%%%%%%%%%%%%
%%%%%%%%%%%%%%%%%%%%%%%%%%%%%%%%%%%%%%%%%%%%%%%%%%%%%%%%%%%%%%%%%%%%%%%%%%%%%%%%%%%%%%%%%
\begin{table*} % [h!]
\small
\begin{center}
\begin{tabular*}{0.60\textwidth}{@{\extracolsep{\fill}}lllll}
\hline
2BDCN/Method     & Basis set   & No.\ functions & $EA$\,(eV) & $EBE$\,(eV) \\
\hline
EOM-CCSD        & DZ          & 176    & -0.113 &   \\
EOM-CCSD        & 6-31G**     & 264    & 0.0677  & \\
% EOM-CCSD        & svp         & 264    & 0.333  & \\
% EOM-CCSD        & dzp         & 280    & 0.271  & \\
% EOM-CCSD        & DZP         & 280    & 0.364  &   \\
% EOM-CCSD        & cc-pVDZ     & 264    & 0.301  & \\
% EOM-CCSD        & cc-pVTZ     & 592   & 0.711     &   \\
EOM-CCSD        & aug-cc-pVDZ & 440    & 0.697 &  3.567 \\
EOM-CCSD(T)     & aug-cc-pVDZ & 440    & 0.697 &    \\
% EOM-CCSD        & aug-cc-pVTZ &        &     &    \\
                &             &     &       &   \\
EOM-CC2         & aug-cc-pVDZ & 440    &  1.033  & \\
% ADC(2)          & aug-cc-pVDZ &        &      &    \\
                &             &     &        &  \\
$\Delta$-MP2       & aug-cc-pVDZ & 440    & 0.0227 &  \\
% $\Delta$-MP2       & 6-311++G(d,p) & 408  & 0.488 &  \\
$\Delta$-CCSD      & aug-cc-pVDZ & 440    & 0.601 & \\
$\Delta$-CC2       & aug-cc-pVDZ & 440    & 0.611 & \\
                &             &     &      &    \\
$\Delta$-DFT/B3LYP & 6-311++G(d,p) & 408 & 1.228 & \\
% \ib{$\Delta$-DFT/B3LYP} & aug-cc-pVTZ &    &   &  \\
% $\Delta$-DFT/B3LYP & 6-31G*  & 256   & 0.736 & \\
% $\Delta$-DFT/B3LYP & 6-31G**     & 280 &  0.744 & \\
% $\Delta$-DFT/B3LYP & 6-31++G(d,p)    & 352    &  1.162      &   \\
% $\Delta$-DFT/B3LYP & cc-pVDZ     &  264      & 0.889     &          \\
% $\Delta$-DFT/B3LYP & cc-pVTZ     &  592      & 1.118     &         \\
% $\Delta$-DFT/B3LYP & svp         &  264      &  0.960    &     \\
                &             &     &         &  \\
$\Delta$-SCF    & aug-cc-pVDZ & 440    & 0.121  &  \\
$\Delta$-SCF    & 6-311++G(d,p) & 408 & 0.117 &   \\
                &             &     &         &  \\
-KS-LUMO        &  aug-cc-pVDZ &  440 & 2.685 & \\
Koopmans theorem & aug-cc-pVDZ       & 440       & -0.617 & \\    
2nd-order pole (97.9\%)  & aug-cc-pVDZ & 440       & -0.171 & \\
3rd-order pole (98.3\%)  & aug-cc-pVDZ & 440       & -0.330 &  \\
OVGF (98.2\% )           & aug-cc-pVDZ & 440       & -0.292 &     \\
\hline
\end{tabular*}
\caption{
The lowest electron affinity (EA) of 4,4'-dicyano-1,1'-biphenyl (2BDCN) 
computed by various methods and the exciton binding energy ($EBE$). 
The geometry of the neutral molecule has been 
optimized at DFT/B3LYP/aug-cc-pVDZ level.}
\label{table-2bdcn}
\end{center}
\end{table*}
%%%%%%%%%%%%%%%%%%%%%%%%%%%%%%%%%%%%%%%%%%%%%%%%%%%%%%%%%%%%%%%%%%%%%%%%%%%%%%%%%%%%%%
%%%%%%%%%%%%%%%%%%%%%%%%%%%%%%%%%%%%%%%%%%%%%%%%%%%%%%%%%%%%%%%%%%%%%%%%%%%%%%%%%%%%%%%%%
\begin{table*} % [h!]
\small
\begin{center}
\begin{tabular*}{0.59\textwidth}{@{\extracolsep{\fill}}lllll}
\hline
BDT/Method     & Basis set   & No.\ functions & $IP$\,(eV) & $EBE$\,(eV)\\
\hline
EOM-CCSD        & DZ          & 108       & 7.492  & \\
EOM-CCSD        & 6-31G**     & 150    & 7.669  & \\
EOM-CCSD        & svp         & 150    & 7.671  & \\
EOM-CCSD        & cc-pVDZ     & 150    & 7.642  & \\
EOM-CCSD        & dzp         & 166    & 7.632  & \\
EOM-CCSD        & cc-pVTZ     & 332    & 7.922  & \\
EOM-CCSD        & aug-cc-pVDZ & 246    & 7.816  & 3.868 \\
EOM-CCSD(T)     & aug-cc-pVDZ & 246    & 7.816  & \\
% EOM-CCSD(T)     & aug-cc-pVTZ & 514    & 7.99356 &  \\
% EOM-CCSD(T)     & aug-cc-pVTZ & 514    & 7.994 &  \\
% EOM-CCSD        & aug-cc-pVTZ & 514    & 7.99356 &  \\
% EOM-CCSD        & aug-cc-pVTZ & 514    & 7.994 &  \\
                &             &     &         & \\
EOM-CCSD(2)     & aug-cc-pVDZ & 246 & 7.883 &   \\
% EOM-CCSD(2)     & aug-cc-pVTZ & 514 & 8.166 &   \\
EOM-CC2         & aug-cc-pVDZ & 246    & 7.497 &   \\
% ADC(2)          & aug-cc-pVDZ & 246    & 9.073 &   \\
                &             &     &        &  \\
$\Delta$-MP2       & aug-cc-pVDZ & 246 & 8.114  & \\
$\Delta$-CCSD      & 6-31G**    & 150 & 7.590 & \\
$\Delta$-CCSD      & aug-c-pVDZ     & 246 & 7.858 & \\
$\Delta$-CC2      & aug-cc-pVDZ     & 246 & 7.874 & \\
                &             &     &        &  \\
$\Delta$-DFT/B3LYP & DZ       &  108  & 7.690  &         \\
$\Delta$-DFT/B3LYP & 6-31G*     & 140  & 7.527  &         \\
$\Delta$-DFT/B3LYP & svp         &  150   & 7.559  &      \\
$\Delta$-DFT/B3LYP & cc-pVDZ     &  150      & 7.549  &           \\
$\Delta$-DFT/B3LYP & 6-31G**     & 158  & 7.531  &         \\
$\Delta$-DFT/B3LYP & 6-311G*     & 196  & 7.659  &    \\
$\Delta$-DFT/B3LYP & 6-31++G(d,p)    & 196  & 7.750  &         \\
$\Delta$-DFT/B3LYP & 6-311+G(d,p)    & 228  & 7.759  &         \\
$\Delta$-DFT/B3LYP & 6-311++G(d,p)    & 234  & 7.770  &         \\
$\Delta$-DFT/B3LYP & cc-pVTZ     &  332      & 7.640    &           \\
$\Delta$-DFT/B3LYP & aug-cc-pVDZ & 246 & 7.669 &   \\
$\Delta$-DFT/B3LYP & aug-cc-pVTZ & 514 & 7.671 &   \\
                &             &     &         &  \\
$\Delta$-SCF    & aug-cc-pVDZ & 246 & 7.149 &  \\
                &             &     &         &  \\
-KS-HOMO         & aug-cc-pVDZ       & 246 & 5.865 & \\
Koopmans theorem & aug-cc-pVDZ       & 246 & 8.038 & \\
2nd-order pole (87.4\%) & aug-cc-pVDZ & 246 &  7.457 & \\
3rd-order pole (90.1\%) & aug-cc-pVDZ & 246 & 7.960  & \\ 
OVGF (89.3\%) & aug-cc-pVDZ & 246 & 7.750 & \\
\hline
\end{tabular*}
\caption{
The lowest ionization energy ($IP$) of benzenedithiol (BDT) 
computed by various methods and the exciton binding energy ($EBE$). 
The geometry of the neutral molecule has been 
optimized at DFT/B3LYP/aug-cc-pVDZ level.}
\label{table-bdt}
\end{center}
\end{table*}
%%%%%%%%%%%%%%%%%%%%%%%%%%%%%%%%%%%%%%%%%%%%%%%%%%%%%%%%%%%%%%%%%%%%%%%%%%%%%%%%%%%%%%
%
%%%%%%%%%%%%%%%%%%%%%%%%%%%%%%%%%%%%%%%%%%%%%%%%%%%%%%%%%%%%%%%%%%%%%%%%%%%%%%%% 
% \begin{figure}[h!]
% $ $\\[6ex]
% \centerline{\hspace*{0.ex}\includegraphics[width=0.4\textwidth,angle=0]{HOMO37_bdt_cfour_augccpvdz.eps}\hspace*{0.ex}}
% $ $\\[-1.4ex]
% \caption{The almost singly occupied natural orbital of the extra hole 
% (``HOMO'') 
% in the BDT cation is delocalized
% over the whole molecule. Result of IP-EOM-CCSD/aug-cc-pVDZ calculations.}
% \label{fig:homo-bdt}
% \end{figure}
%%%%%%%%%%%%%%%%%%%%%%%%%%%%%%%%%%%%%%%%%%%%%%%%%%%%%%%%%%%%%%%%%%%%%%%%%%%%%%%%
%
%%%%%%%%%%%%%%%%%%%%%%%%%%%%%%%%%%%%%%%%%%%%%%%%%%%%%%%%%%%%%%%%%%%%%%%%%%%%%%%%%%%%%%%%%
\begin{table*} % [h!]
\small
\begin{center}
\begin{tabular*}{0.59\textwidth}{@{\extracolsep{\fill}}lllll}
\hline
2BDT/Method                         & Basis set   & No.\ functions & $IP$\,(eV) & $EBE$\,(eV) \\
\hline
EOM-CCSD                            & DZ          & 176    & 7.294 & \\
EOM-CCSD                            & 6-31G**     & 254    & 7.340  &  \\
EOM-CCSD                            & svp         & 254    & 7.483  & \\
EOM-CCSD                            & dzp         & 276    & 7.437  & \\
EOM-CCSD                            & cc-pVDZ     & 254    & 7.433  & \\
% \ib{EOM-CCSD}                           & cc-pVTZ     &        &        & \\
EOM-CCSD                            & aug-cc-pVDZ & 420    & 7.593  & 3.470 \\
EOM-CCSD(T)                         & aug-cc-pVDZ & 420    & 7.593  & \\
% \ib{EOM-CCSD}                       & aug-cc-pVTZ &        &        & \\
                &             &     &          & \\
EOM-CCSD(2)                             & aug-cc-pVDZ & 420    & 7.747 & \\
EOM-CC2                             & aug-cc-pVDZ & 420    & 7.303  &  \\
% ADC(2)                              & aug-cc-pVDZ &        &        &  \\
                &             &     &        &  \\
$\Delta$-MP2                       & 6-31G*       & 254 &  8.629  & \\
$\Delta$-MP2                       & 6-311++G(d,p) & 394 &  8.831  & \\
$\Delta$-CCSD                      & 6-31G*       & 254 &  7.436  & \\
$\Delta$-CCSD                      & aug-cc-pVDZ   & 420 & 7.714    & \\
$\Delta$-CC2                      & aug-cc-pVDZ   & 420 & 7.888 & \\
                &             &     &        &  \\
$\Delta$-DFT/B3LYP & DZ            & 176    & 7.255    &      \\
$\Delta$-DFT/B3LYP & 6-31G*     & 238 &  7.092 &         \\
$\Delta$-DFT/B3LYP & svp           & 254   & 7.172    &      \\
$\Delta$-DFT/B3LYP & cc-pVDZ       & 254    & 7.149 &           \\
$\Delta$-DFT/B3LYP & 6-31G**     & 268 &  7.097 &         \\
$\Delta$-DFT/B3LYP & 6-311G*       & 298    & 7.248   &       \\
% $\Delta$-DFT/B3LYP/opt 6-31+G(d,p)  & 6-31+G(d,p) & 334 & 7.292 & \\
% $\Delta$-DFT/B3LYP/opt 6-31+G(d,p)  & 6-311++G(d,p) & 394 & 7.288 & \\
$\Delta$-DFT/B3LYP & 6-311G**      & 328    & 7.256   &       \\
$\Delta$-DFT/B3LYP & cc-pVTZ       & 568    & 7.242    &           \\
% \ib{$\Delta$-DFT/B3LYP} & aug-cc-pVTZ   & 882    &   & \\
$\Delta$-DFT/B3LYP & aug-cc-pVDZ   & 420    & 7.269       &  \\
% \ib{$\Delta$-DFT/B3LYP} & aug-cc-pVTZ   & 882    & 2BDT\_AugCCpVTZ\_dft\_optimized\_6311++g\_dp\_C2\_n\_IP\_EA\_tight   & hitchcock \\
                &             &     &       &   \\
$\Delta$-SCF                        & aug-cc-pVDZ & 420 & 6.671  &       \\
$\Delta$-SCF                        & 6311++G(d,p) & 394 & 6.819  &       \\
                &             &     &       &   \\
-KS-HOMO                           & 6-311++G(d,p) & 394    & 5.825 & \\
Koopmans theorem                   & 6-311++G(d,p) & 394    & 7.778  & \\    
2nd-order pole (86.5\%)            & 6-311++G(d,p) & 394    & 7.260 & \\
3rd-order pole (89.8\%)            & 6-311++G(d,p) & 394    & 7.629  &  \\
OVGF (88.8\% )                     & 6-311++G(d,p) & 394    & 7.522  &     \\
\hline
\end{tabular*}
\caption{The lowest ionization energy ($IP$) of dibenzenedithiol (2BDT) 
computed by various methods and the exciton binding energy ($EBE$). 
The geometry of the neutral molecule has been 
optimized at DFT/B3LYP/6-31++G(d,p) level \cite{why-not-aug-cc-pvdz}.}
\label{table-2bdt}
\end{center}
\end{table*}
%%%%%%%%%%%%%%%%%%%%%%%%%%%%%%%%%%%%%%%%%%%%%%%%%%%%%%%%%%%%%%%%%%%%%%%%%%%%%%%%%%%%%%
%%%%%%%%%%%%%%%%%%%%%%%%%%%%%%%%%%%%%%%%%%%%%%%%%%%%%%%%%%%%%%%%%%%%%%%%%%%%%%%%%%%%%%%%%
\begin{table*} % [h!]
\small
\begin{center}
\begin{tabular*}{0.58\textwidth}{@{\extracolsep{\fill}}lllll}
\hline
C6MT/Method     & Basis set   & No.\ functions & $IP$\,(eV) & $EBE$\,(eV) \\
\hline
EOM-CCSD        & DZ          & 106       & 8.543 & \\
EOM-CCSD        & 6-31G**     & 172    & 8.741 & \\
EOM-CCSD        & svp         & 172    & 8.806 &  \\
EOM-CCSD        & dzp         & 183    & 8.785 &  \\
EOM-CCSD        & cc-pVDZ     & 172    & 8.803 &  \\
EOM-CCSD        & cc-pVTZ     & 291    & 9.107 &  \\
EOM-CCSD        & aug-cc-pVDZ & 291    & 8.966 &  4.130 \\
EOM-CCSD(T)     & aug-cc-pVDZ & 291    & 8.966 &    \\
% EOM-CCSD(T)     & aug-cc-pVDZ & c6t\_ccsdt\_augccpvdz\_cation.out       &       &  \\
% EOM-CCSD        & aug-cc-pVTZ &        &       &  \\
                &             &     &          & \\
EOM-CCSD(2)     & aug-cc-pVDZ & 291    & 8.799 & \\
% EOM-CCSD(2)     & aug-cc-pVTZ & 648    & 9.072 & \\
EOM-CC2         & aug-cc-pVDZ & 291  &  8.806   &     \\
ADC(2)          & aug-cc-pVDZ & 291  &  8.623   &  \\
                &             &     &         & \\
$\Delta$-MP2       & aug-cc-pVDZ & 291    & 9.0448 & \\
$\Delta$-CCSD & aug-cc-pVDZ & 291    & 8.922  & \\
$\Delta$-CC2      & aug-cc-pVDZ & 291 & 9.012 & \\
                &             &     &        &  \\
$\Delta$-DFT/B3LYP & DZ         & 106   & 8.989 & \\
$\Delta$-DFT/B3LYP & svp         & 172    & 8.925 &       \\
$\Delta$-DFT/B3LYP & 6-31G*     & 137    & 8.930 &         \\
$\Delta$-DFT/B3LYP & cc-pVDZ     & 172  & 8.918   &            \\
$\Delta$-DFT/B3LYP & 6-31G**    & 179    & 8.926 & \\
$\Delta$-DFT/B3LYP & 6-311G*    & 176   & 9.019 &      \\
$\Delta$-DFT/B3LYP & cc-pVTZ     & 410  & 8.975    &            \\
$\Delta$-DFT/B3LYP & aug-cc-pVDZ & 291 & 8.994  &  \\
$\Delta$-DFT/B3LYP & aug-cc-pVTZ & 648 & 8.990 & \\
                &             &     &         & \\
$\Delta$-SCF    & aug-cc-pVDZ    & 291  & 7.988 &  \\
                &             &     &  & \\
-KS-HOMO         & aug-cc-pVDZ       & 291    &  6.490 & \\    
Koopmans theorem & aug-cc-pVDZ       & 291    &  9.575 & \\    
2nd-order pole (90.6\%)  & aug-cc-pVDZ & 291    & 8.632 & \\
3rd-order pole (91.1\%)  & aug-cc-pVDZ & 291    & 9.109 &   \\
OVGF (90.7\% )           & aug-cc-pVDZ & 291    & 9.058 &      \\
\hline
\end{tabular*}
\caption{
The lowest ionization energy ($IP$) of hexanemonothiol (C6MT)
computed by various methods and the exciton binding energy ($EBE$). 
The geometry of the neutral molecule has been 
optimized at DFT/B3LYP/aug-cc-pVDZ level.}
\label{table-c6t}
\end{center}
\end{table*}
%%%%%%%%%%%%%%%%%%%%%%%%%%%%%%%%%%%%%%%%%%%%%%%%%%%%%%%%%%%%%%%%%%%%%%%%%%%%%%%%%%%%%%
%%%%%%%%%%%%%%%%%%%%%%%%%%%%%%%%%%%%%%%%%%%%%%%%%%%%%%%%%%%%%%%%%%%%%%%%%%%%%%%%%%%%%%%%%
\begin{table*} % [h!]
\small
\begin{center}
\begin{tabular*}{0.58\textwidth}{@{\extracolsep{\fill}}lllll}
\hline
C6DT/Method     & Basis set   & No.\ functions & $IP$\,(eV) & $EBE$\,(eV) \\
\hline
EOM-CCSD        & DZ          & 124    & 8.591 & \\
EOM-CCSD        & 6-31G*      & 148    & 8.748 & \\
EOM-CCSD        & svp         & 190    & 8.847     &   \\
EOM-CCSD        & dzp         & 206    & 8.829     & \\
EOM-CCSD        & cc-pVDZ     & 190    & 8.846  & \\
EOM-CCSD        & cc-pVTZ     & 444    & 9.162 & \\
EOM-CCSD        & aug-cc-pVDZ & 318    & 9.026  & 4.129 \\
EOM-CCSD(T)     & aug-cc-pVDZ & 318    & 9.026  & \\
% EOM-CCSD        & aug-cc-pVTZ &        &        & \\
                &             &     &        &  \\
EOM-CCSD(2)         & aug-cc-pVDZ & 318    & 8.859 & \\
EOM-CC2         & aug-cc-pVDZ & 318    & 8.482     &    \\
ADC(2)          & aug-cc-pVDZ & 318    & 8.696     &    \\
                &             &     &         & \\
$\Delta$-MP2       & aug-cc-pVDZ & 318    & 9.092 &        \\
$\Delta$-CCSD      & aug-cc-pVDZ & 318    & 8.974 &   \\
$\Delta$-CC2      & aug-cc-pVDZ & 318    & 9.063       & \\
                &             &     &          & \\
$\Delta$-DFT/B3LYP & DZ          & 124  & 8.209    &      \\
$\Delta$-DFT/B3LYP & svp         & 190  & 8.111    &      \\
$\Delta$-DFT/B3LYP & cc-pVDZ     & 190  & 8.136 &             \\
$\Delta$-DFT/B3LYP & 6-31G**     & 198  & 8.167   &           \\
$\Delta$-DFT/B3LYP & 6-311G**    & 244  & 8.287  &      \\
$\Delta$-DFT/B3LYP & cc-pVTZ     & 444    & 8.246  &            \\
$\Delta$-DFT/B3LYP & aug-cc-pVDZ & 318 & 8.273 &   \\
$\Delta$-DFT/B3LYP & aug-cc-pVTZ & 698 & 8.279   & \\
                &             &     &        &  \\
$\Delta$-SCF    & aug-cc-pVDZ & 318    & 8.042 & \\
                &             &     &        &  \\
-KS-HOMO         & aug-cc-pVDZ       & 318    &  6.535 & \\    
Koopmans theorem & aug-cc-pVDZ       & 318    & 9.628 & \\    
2nd-order pole (90.7\%)  & aug-cc-pVDZ & 318    & 8.712 & \\
3rd-order pole (91.1\%)  & aug-cc-pVDZ & 318    & 9.158 &   \\
OVGF (90.7\% )           & aug-cc-pVDZ & 318    & 9.116 &      \\
\hline
\end{tabular*}
\caption{
The lowest ionization energy ($IP$) of hexanedithiol (C6DT)
computed by various methods and the exciton binding energy ($EBE$). 
The geometry of the neutral molecule has been 
optimized at DFT/B3LYP/aug-cc-pVDZ level.}
\label{table-c6dt}
\end{center}
\end{table*}
%%%%%%%%%%%%%%%%%%%%%%%%%%%%%%%%%%%%%%%%%%%%%%%%%%%%%%%%%%%%%%%%%%%%%%%%%%%%%%%%%%%%%%
%%%%%%%%%%%%%%%%%%%%%%%%%%%%%%%%%%%%%%%%%%%%%%%%%%%%%%%%%%%%%%%%%%%%%%%%%%%%%%%% 
% \begin{figure}[h!]
% $ $\\[6ex]
%%% \centerline{\hspace*{0.ex}\includegraphics[width=0.4\textwidth,angle=0]{HOMO_c6dt_cfour_augccpvdz.eps}\hspace*{0.ex}}
% \centerline{\hspace*{0.ex}\includegraphics[width=0.4\textwidth,angle=0]{HOMO_c6dt_cfour_ccpvdz.eps}\hspace*{0.ex}}
% $ $\\[-1.4ex]
% \caption{The almost singly occupied natural orbital of the extra hole 
% (``HOMO'') 
% in the 
% hexanedithiol cation is localized on the sulfur atoms at the two molecular ends.
% Result of IP-EOM-CCSD/aug-cc-pVDZ calculations.}
% \label{fig:homo-c6dt}
% \end{figure}
%%%%%%%%%%%%%%%%%%%%%%%%%%%%%%%%%%%%%%%%%%%%%%%%%%%%%%%%%%%%%%%%%%%%%%%%%%%%%%%

%%%%%%%%%%%%%%%%%%%%%%%%%%%%%%%%%%%%%%%%%%%%%%%%%%%%%%%%%%%%%%%%%%%%%%%%%%%%%%%%%%%%%%%%%
\begin{table*} % [h!]
\small
\begin{center}
\begin{tabular*}{0.91\textwidth}{@{\extracolsep{\fill}}lllll}
\hline
Molecule/Optimization & Method     & Basis set   & No.\ functions & IP\,(eV) \\
\hline
BDT &                 &             &        &         \\
optimized B3LYP/aug-cc-pVDZ    & 2nd-order pole (87.4\%) & aug-cc-pVDZ & 246 & 7.457 \\
    & 3rd-order pole (90.1\%) & aug-cc-pVDZ & 246 & 7.960  \\ 
    & OVGF (89.3\%) & aug-cc-pVDZ & 246 & 7.750  \\
 &                 &             &        &         \\
% & $\Delta$-MP2       & STO-6G  & 54     & 5.020   \\
 & 2nd-order pole (91.9\%)  & STO-6G & 54 & 4.702 \\
 & 3rd-order pole (91.5\%)  & STO-6G & 54 & 4.775 \\
 & OVGF (91.4\%) & STO-6G & 54 & 4.754 \\
\hline
\ce{S}-\ce{C6H4}-\ce{S}                      &              &        &    &       \\
Ref.~\cite{Shimazaki:09}      &   MP2-based   & STO-6G  & 52   & 6.5267 \\
optimized B3LYP/aug-cc-pVDZ 
& EOM-CCSD       & aug-cc-pVDZ & 228 & 8.920 \\
& 2nd-order pole (87.3\%) & STO-6G & 52 & 5.139 \\
& 3rd-order pole (83.6\%) & STO-6G & 52 & 5.319\\
& OVGF (82.9\%) & STO-6G & 52 & 5.353 \\
& & & & \\
optimized MP2/STO-6G 
% & $\Delta$-MP2 & STO-6G & 52 & 4.528 \\ 
 & 2nd-order pole (88.3\%) & STO-6G & 52 & 5.382 \\
 & 3rd-order pole (85.0\%) & STO-6G & 52 & 5.499 \\
 & OVGF (84.4\%) & STO-6G & 52 & 5.521\\
\hline
\ce{S}-\ce{C6H3F}-\ce{S}  &  & & &  \\
Ref.~\cite{Shimazaki:09} & MP2-based & STO-6G & 56 & 6.8019 \\
optimized B3LYP/aug-cc-pVDZ 
% & $\Delta$-MP2 & STO-6G & 56 & 4.713 \\
& EOM-CCSD       & aug-cc-pVDZ & 242 & 8.229 \\
& 2nd-order pole (87.4\%) & STO-6G & 56 & 5.235 \\
& 3rd-order pole (83.7\%) & STO-6G & 56 & 5.370\\
& OVGF (83.0\%) & STO-6G & 56 & 5.395 \\
& & & & \\
optimized MP2/STO-6G 
 & 2nd-order pole (88.2\%) & STO-6G & 56 & 5.444 \\
 & 3rd-order pole (85.1\%) & STO-6G & 56 & 5.522 \\
 & OVGF (84.5\%) & STO-6G & 56 & 5.534 \\
\hline
\end{tabular*}
\caption{MP2-based results for the lowest ionization energy ($IP$)
of the benzenedithiol (BDT) and BDT-like molecules considered in ref.~\cite{Shimazaki:09} along with results obtained within methods 
utilized in this paper.}
\label{table-bdt-mp2}
\end{center}
\end{table*}
%%%%%%%%%%%%%%%%%%%%%%%%%%%%%%%%%%%%%%%%%%%%%%%%%%%%%%%%%%%%%%%%%%%%%%%%%%%%%%%%%%%%%%
%%%%%%%%%%%%%%%%%%%%%%%%%%%%%%%%%%%%%%%%%%%%%%%%%%%%%%%%%%%%%%%%%%%%%%%%%%%%%%%%%%%%%%%%%
\begin{table*} % [h!]
\small
\begin{center}
% \begin{tabular*}{0.93\textwidth}{@{\extracolsep{\fill}}lllll}
\begin{tabular*}{0.6\textwidth}{@{\extracolsep{\fill}}llll}
\hline
% Method & Molecule & $IP$\,(eV) & $EA$\,(eV) & HOMO-LUMO gap\,(eV) \\
Method & $IP$\,(eV) & $EA$\,(eV) & HOMO-LUMO gap\,(eV) \\
\hline
% BDT    & 
EOM-CCSD & 7.492 & -2.016 & 9.500 \\
%       & 
OVGF     & 7.560 & -2.073 & 9.633 \\
%       & 
$GW$     & 6.9   & -2.2   & 9.1   \\ 
% \ce{C6H4S2} & EOM-CCSD & 8.229 & 1.617  & 7.213 \\
%             & OVGF     & 8.523 & 1.763  & 6.760 \\
%             & $GW$     & 7.9   &  2.3   & 5.6   \\ 
\hline
\end{tabular*}
\caption{Results for the lowest ionization energy ($IP$), electron affinity ($EA$), and HOMO-LUMO gap obtained by using several methods and the same double zeta (DZ) basis sets. 
Notice that the $GW$ estimates taken from ref.~\cite{Thygesen:11c} significantly differ from the EOM-CCSD values, which are very accurately estimated within the OVGF. One should remark that the fact that all methods predict a stable anion ($EA>0$) is an artefact of the small (DZ) basis set utilized in this table.}
\label{table-bdt-gw}
\end{center}
\end{table*}
%%%%%%%%%%%%%%%%%%%%%%%%%%%%%%%%%%%%%%%%%%%%%%%%%%%%%%%%%%%%%%%%%%%%%%%%%%%%%%%%%%%%%%
%%%%%%%%%%%%%%%%%%%%%%%%%%%%%%%%%%%%%%%%%%%%%%%%%%%%%%%%%%%%%%%%%%%%%%%%%%%%%%%% 
% \begin{figure}[h!]
% $ $\\[6ex]
% \centerline{\hspace*{0.ex}\includegraphics[width=0.4\textwidth,angle=0]{HF_HOMO_1BDT2Au_AugCCpVDZ.eps}\hspace*{0.ex}}
% $ $\\[2ex]% \centerline{\hspace*{0.ex}\includegraphics[width=0.4\textwidth,angle=0]{HF_HOMO_C6MT2Au_AugCCpVDZ.eps}\hspace*{0.ex}}
% $ $\\[2ex]% \centerline{\hspace*{0.ex}\includegraphics[width=0.4\textwidth,angle=0]{HF_HOMO_C6DT2Au_AugCCpVDZ.eps}\hspace*{0.ex}}
% $ $\\[-1.4ex]
% \caption{HOMO distributions for molecules linked to two gold atoms (Au-BDT-Au, C6MT-Au, and Au-C6DT-Au).}
% \label{fig:homo-molec-2Au}
% \end{figure}
%%%%%%%%%%%%%%%%%%%%%%%%%%%%%%%%%%%%%%%%%%%%%%%%%%%%%%%%%%%%%%%%%%%%%%%%%%%%%%%%

\providecommand*{\mcitethebibliography}{\thebibliography}
\csname @ifundefined\endcsname{endmcitethebibliography}
{\let\endmcitethebibliography\endthebibliography}{}

%%%%%%%%%%%%%%%%%%%%%%%%%%%%%%%%%%%%%%%%%%%%%%%%%%%%%%%%%%%%%%%%%%%%%%%%%%%%%%%%%%%%%%
\end{document}